\documentclass[journal,twoside, print]{ieeecolor}
\usepackage{tmi}
\usepackage{upgreek}
\usepackage{cite}
\usepackage{amsmath,amssymb,amsfonts}
\usepackage{algorithmic}
\usepackage{graphicx}
\usepackage{textcomp}
\usepackage{marvosym}

\def\BibTeX{{\rm B\kern-.05em{\sc i\kern-.025em b}\kern-.08em
    T\kern-.1667em\lower.7ex\hbox{E}\kern-.125emX}}
\begin{document}
\title{From Pixel to Slide image:  Polarization Modality-based Pathological Diagnosis Using Representation Learning}
\author{Jia Dong*, Yao Yao*, Yang Dong, and Hui Ma\textsuperscript{\Letter}
\thanks{Both authors contributed equally: Jia Dong; Yao Yao.}
\thanks{Corresponding author: Hui Ma (mahui@tsinghua.edu.cn).}
\thanks{Jia Dong is currently with Department of Statistical Science at University College London (jia.dong.23@ucl.ac.uk).}
\thanks{Jia Dong, Yang Dong, and Hui Ma are with Guangdong
Engineering Center of Polarization Imaging and Sensing Technology,
Shenzhen Key Lab for Minimal Invasive Medical Technologies, Tsinghua
Shenzhen International Graduate School, Tsinghua University, Shenzhen, 518055, China.}
\thanks{Yao Yao, Yang Dong, and Hui Ma are with the Tsinghua Berkeley Shenzhen Institute, Tsinghua Shenzhen International Graduate School, Tsinghua University, Shenzhen, 518055, China.}
\thanks{This work was supported in part by National Natural Science
Foundation of China (NSFC) (Grant Nos. 61527826 and 11974206)
and Shenzhen Bureau of Science and Innovation (Grant No.
JCYJ20170412170814624). }
}

\maketitle

\begin{abstract}
Thyroid cancer is the most common endocrine malignancy, and accurately distinguishing between benign and malignant thyroid tumors is crucial for developing effective treatment plans in clinical practice. Pathologically, thyroid tumors pose diagnostic challenges due to improper specimen sampling. In this study, we have designed a three-stage model using representation learning to integrate pixel-level and slice-level annotations for distinguishing thyroid tumors. This structure includes a pathology structure recognition method to predict structures related to thyroid tumors, an encoder-decoder network to extract pixel-level annotation information by learning the feature representations of image blocks, and an attention-based learning mechanism for the final classification task. This mechanism learns the importance of different image blocks in a pathological region, globally considering the information from each block. In the third stage, all information from the image blocks in a region is aggregated using attention mechanisms, followed by classification to determine the category of the region. Experimental results demonstrate that our proposed method can predict microscopic structures more accurately. After color-coding, the method achieves results on unstained pathology slides that approximate the quality of Hematoxylin and eosin staining, reducing the need for stained pathology slides. Furthermore, by leveraging the concept of indirect measurement and extracting polarized features from structures correlated with lesions, the proposed method can also classify samples where membrane structures cannot be obtained through sampling, providing a potential objective and highly accurate indirect diagnostic technique for thyroid tumors.
\end{abstract}


\section{Introduction}
\label{sec:introduction}

\IEEEPARstart{T}{he} thyroid gland is a crucial organ that regulates various bodily functions such as metabolic rate, energy expenditure, and the function of organs like the heart and brain \cite{1}. Thyroid diseases are among the most common ailments, with thyroid cancer ranking 9th globally in the incidence of all cancers, totaling 586,000 cases in 2022 \cite{2}. Currently, the World Health Organization classification system is primarily used, categorizing thyroid tumors into benign tumors (adenomas), low-risk tumors, and malignant tumors (carcinomas) \cite{3}. Thyroid cancer is the most common endocrine malignancy, accounting for 2.1\% of all new diagnoses of malignancies (excluding skin cancers and in situ carcinomas) annually. The incidence of thyroid cancer has been steadily rising over the past few decades, and if the current trend continues, it might become the fourth most common cancer by 2030 \cite{4}. Thyroid cancer diagnosis involves intricate steps like tissue preparation, histological observation, and immunohistochemistry \cite{5,6,7}. Skilled physicians perform these, aligning with clinical data for precision. Crucially, cancer classification aids tailored treatment plans, prognosis prediction, and treatment evaluation. Different thyroid cancer types necessitate diverse treatments—papillary carcinoma often involves surgery and hormone replacement, while follicular carcinoma may require surgery and radiation \cite{8}. Accurate classification, pivotal for clinical decisions, provides insights into biological traits and informs novel treatments and diagnostics. Thus, precise thyroid cancer classification is fundamental for clinical care, prognosis, and research \cite{9,10}.

Artificial intelligence (AI) technology has the capability to automatically analyze and diagnose pathological images, enhancing the accuracy and efficiency of digital pathology and providing doctors with a more comprehensive and objective reference for diagnosis \cite{11,12,13,14,dong2023polarization}. In the pathological assisted diagnosis of thyroid tumors, there has been extensive research on using whole-slide pathology images \cite{15,16} for the classification of thyroid follicular carcinoma. For example, Wang et al. \cite{17}  developed an algorithm that distinguishes thyroid cancer through three stages: image preprocessing and segmentation, feature extraction, and model prediction. Dov et al. \cite{18}  employed a deep learning algorithm based on two cascaded convolutional neural networks (CNNs) to classify thyroid cancer in whole-slide imaging (WSI) cytological pathology images. The algorithm, trained on 799 slices, demonstrated an area under the Receiver Operating Characteristic (ROC) curve of 0.932 in experimental results. Elliott Range et al. \cite{19} constructed a neural network for predicting malignant tumors in thyroid pathology based on two CNNs. Utilizing 908 fine-needle aspiration biopsy images from 659 patients, the algorithm achieved a sensitivity of 92.0\% and specificity of 90.5\%. Hossiny et al. \cite{20}  employed a cascaded CNN technique to divide the thyroid classification process into two stages, reducing the number of classes at each stage. They further subclassified follicular carcinoma from the first stage into four subtypes, achieving an overall accuracy of 94.69\%. Deep learning has been successfully applied to the cell classification of different types of thyroid tumors \cite{21} and has demonstrated the ability to differentiate between benign and malignant cell samples \cite{22}.

The polarization measurement and imaging method based on Mueller matrix, as an emerging biomedical imaging technique, has demonstrated potential diagnostic capabilities in the detection of various cancer tissues \cite{23,24,25,26}. In recent years, many studies have analyzed the polarization parameters of pathological slices, utilizing them as two-dimensional images for feature extraction and assisting in pathological diagnosis \cite{27,28}. We can summarize the research progress of two main polarization image analysis methods: quantitative feature extraction and deep learning modeling. In the quantitative feature extraction method, the Mueller matrix of pathological samples is first measured, obtaining two-dimensional images of polarization parameters from it \cite{29,30}. Subsequently, through further statistical feature analysis or image texture feature analysis of these two-dimensional images, a set of quantitative parameters that effectively differentiate different types of lesion structures is derived. These parameters provide crucial information about the nature and pathological status of the lesions. On the other hand, deep learning modeling methods involve training deep learning models to automatically learn and extract key features from images, aiming for more accurate classification and diagnosis of pathological structures \cite{31,32}. The training process of deep learning models effectively discovers and utilizes hidden features in images, enhancing the accuracy of analysis and recognition capabilities for pathological structures.

The fusion of polarization imaging and Convolutional Neural Networks (CNN) proves feasible in early clinical trials. Yao et al. \cite{33} explored optimal CNN models for diverse polarization images. In endometrial sample differentiation, CNN1 (AlexNet-based) and CNN2 (ResNet-based) efficiently extracted features from 2D polarization parameter images, achieving reliable classification. Linear polarization and extinction angle parameters yielded the same highest accuracy of 0.870. Zhao et al. \cite{34} proposed a method for giant cell tumor detection, employing a CNN for deep feature extraction and a multi-parameter fusion network for improved accuracy. Xia et al. \cite{35} enhanced breast cancer cell classification using ReSENet, showing superior accuracy by capturing polarization image features. Ma et al. \cite{36} introduced MuellerNet, a 3D-2D hybrid CNN for breast cancer cell classification. This innovative approach achieved higher accuracy by integrating information from polarization and optical intensity images. These studies highlight the promising role of polarization imaging with CNN in accurate pathological analysis.

This article presents a three-stage model for thyroid tumor classification using polarization features. The model integrates pixel-level and slice-level annotations through a representation learning \cite{37} approach. The structure includes a pathology structure recognition method for predicting thyroid tumor-related structures—cells, fibers, and glial structures. An encoder-decoder network distills pixel-level annotation information by learning feature representations of image blocks, connecting annotations at different levels. The features of all image blocks in a region are aggregated using an attention mechanism to form a region-level feature representation, which is then classified by a classifier to determine the region's category. To validate confidence learning, a portion of the pathologist-provided annotations is randomly shuffled, and the model predicts pixels in the unlabeled areas under various levels of artificial noise. Experimental results indicate improved performance with the introduced noise. The three recognition results are visualized in an H\&E color-matched format, presenting the polarized virtual staining in a pathologist-friendly manner for convenient pathological observation. The importance of different polarization parameters during the recognition process is analyzed to understand each parameter's significance. For evaluating the restoration performance of structural information, quantitative metrics, including accuracy and loss functions, demonstrate the model's quick convergence and high consistency in predicting structural information at both stages. The second stage effectively reconstructs the structural information from the first stage, providing an approximation of pathological color images. ROC curves from the proposed network's classification illustrate effective differentiation among malignant thyroid tumors, benign tumors, and tumors with indeterminate malignant potential during the classification process. Notably, the discrimination of malignant thyroid tumors performs exceptionally well. In the third stage, the visual representation of the aggregated region-level features, learned using attention mechanisms, demonstrates the method's feasibility, with features possessing classification capabilities and excellent distinctiveness between regions.

\section{Methods}
\subsection{Pathological Samples}
\begin{figure}[!t]
\centerline{\includegraphics[width=\columnwidth]{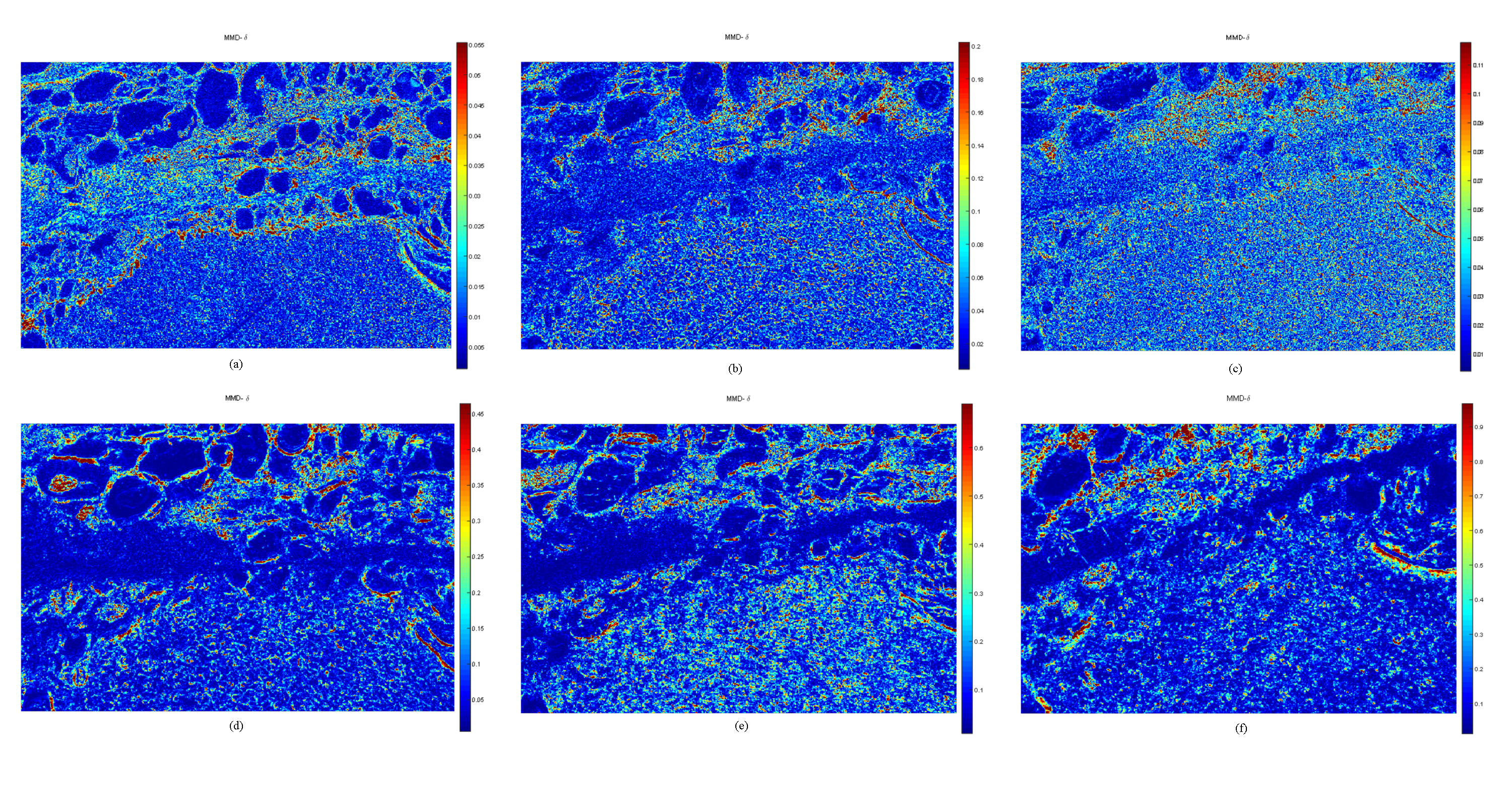}}
\caption{Polarization parameter maps for different slice thicknesses. (a) 4$\upmu$m thick H\&E stained slice. (b) 4$\upmu$m thick unstained slice. (c) 6$\upmu$m thick unstained slice. (d) 8$\upmu$m thick unstained slice. (e) 10$\upmu$m thick unstained slice. (f) 12$\upmu$m thick unstained slice.}
\label{fig1}
\end{figure}
\begin{figure*}[h]
\centerline{\includegraphics[width=\linewidth]{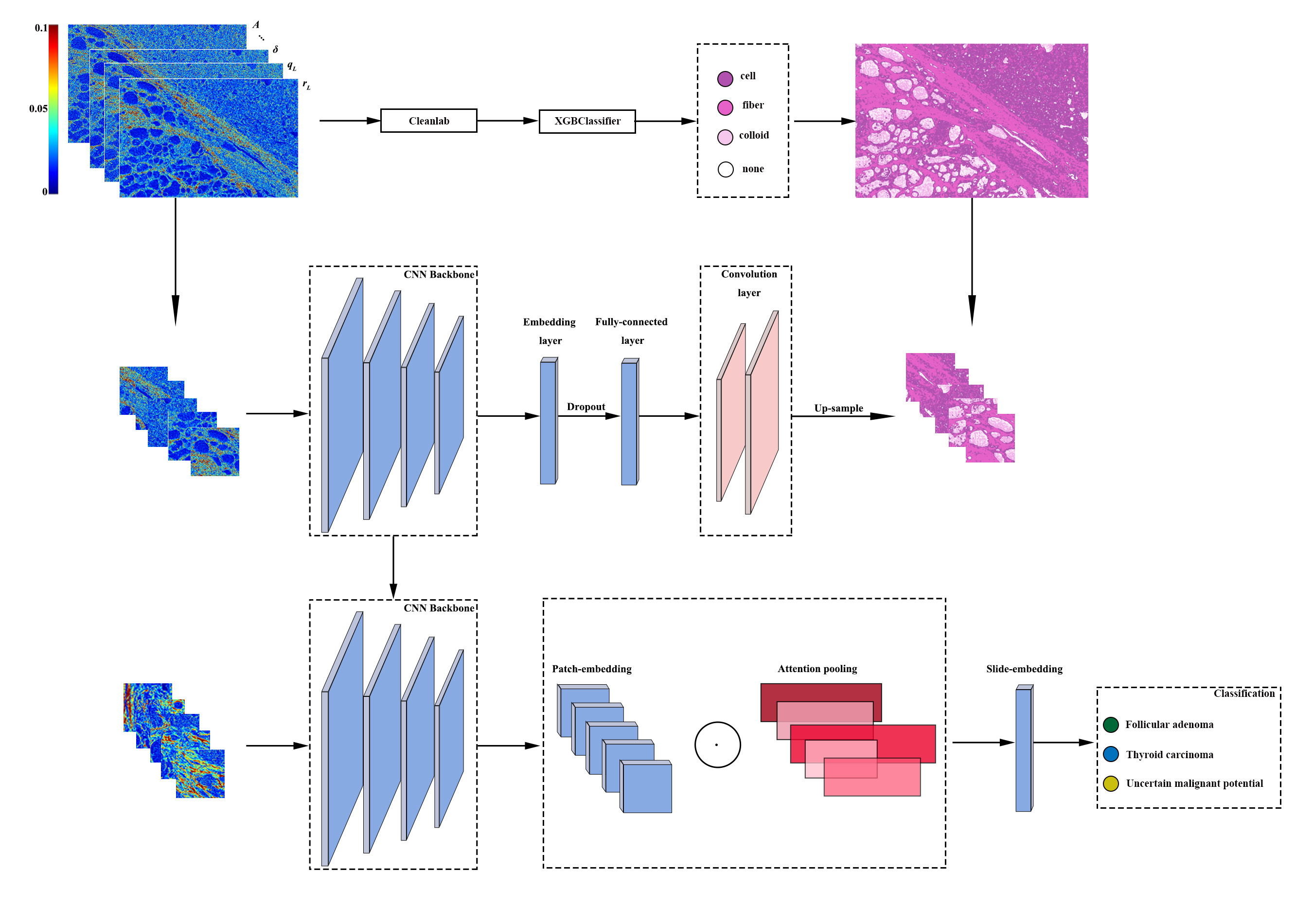}}
\caption{The network architecture for representation learning is divided into three stages: the first stage is microstructure recognition, the second stage involves feature extraction, and the third stage focuses on thyroid tumor classification.}
\label{fig2}
\end{figure*}
In pathology, the preparation of stained slices involves the use of chemical reagents containing harmful substances, demanding strict workflows and staining conditions. Unstained pathological slices allow the observation of cell and tissue morphology without the need for staining, eliminating the time and cost associated with the staining process. This presents advantages for rapid testing and initial screening. Leveraging the clinical significance of unstained pathological slices, we explore the use of polarization information to classify thyroid tumors. To investigate the impact of unstained slice thickness on polarization imaging quality, we conducted experiments using multiple consecutive slices from the same wax block, with thicknesses ranging from 4$\upmu$m to 12$\upmu$m. One 4$\upmu$m thick slice underwent routine H\&E staining to create a stained pathological slice, while the others (4$\upmu$m, 6$\upmu$m, 8$\upmu$m, 10$\upmu$m, 12$\upmu$m) remained unstained after deparaffinization and were used to create unstained pathological slices. All six slices were subjected to polarization imaging using a Mueller matrix microscope, and the polarization parameter images were compared. The results, as shown in Fig. 1, indicate that the 12$\upmu$m thick unstained slice exhibits high polarization signal contrast and a clean background, resembling the polarization parameter image of the H\&E stained slice. Consequently, the 12$\upmu$m thick unstained slice was selected for subsequent experiments. Considering the pathological significance and the exploration of the impact of unstained slice thickness, we selected 116 slices from patients who underwent radical tumor resection at the General Hospital of Southern Medical University. These samples included 53 cases of confirmed malignant thyroid cancer, 48 cases of benign thyroid adenoma, and 35 cases of tumors with indeterminate malignant potential. Each postoperative specimen underwent routine sectioning, and for each case, one sample from the tumor's interior without necrosis was chosen. According to pathology sampling standards, each tissue block was approximately (1.5-2) × 1 × 0.2cm³. Two consecutive slices, one 4$\upmu$m thick stained with H\&E for high-resolution full-slide color pathology imaging, and the other 12$\upmu$m thick unstained, were used for full polarization imaging.

\subsection{Algorithm Architecture}
The framework consists of a label classifier model for pathological structure recognition, a PBP analyzing system for illustrating the polarization features of target microstructures in thyroid tumors, and a CNN for Thyroid carcinoma classification, as shown in Fig. 2. This framework uses polarization information to quantitatively diagnose thyroid tumors, where architecture and parameters are illustrated as follows.

\subsubsection{Confidence Learning-Based Pixel-Level Pathological Structure Recognition}
Confidence learning \cite{38} is a weakly supervised learning approach that focuses on the quality (confidence) of labels. It aims to enhance the quality of data labels by characterizing and identifying erroneous labels in the dataset. This method ensures that the model's performance remains unaffected by noisy labels, ultimately improving the accuracy of the trained model. The key components of confidence learning include:
\begin{enumerate}
    \item Estimation of the joint distribution of noisy labels and true labels.
    \item Identification and removal of low-confidence samples.
    \item Retraining the model after eliminating incorrectly labeled samples.
\end{enumerate}

This approach is particularly useful in addressing issues such as errors in data annotation or inconsistency in labeling standards encountered in practical scenarios. To better handle label noise introduced during the manual annotation process, we employed confidence learning to prune the annotations of the original microscopic structures. The confidence learning process involves three steps:

\begin{enumerate}
    \item Estimating the joint distribution of noisy labels $\tilde{y}$ and true labels $y^*$ to describe the noise in labels for different categories.
    \item Finding and filtering out low-confidence annotations.
    \item Reassigning weights to different categories and retraining the model on the remaining clean data.
\end{enumerate}

The dataset $X$ for thyroid microscopic structures is represented as $(x, \tilde{y})_n$, consisting of $n$ samples with $m$ classes of noisy labels $\tilde{y}$. Cross-validation is performed using a third-party model to calculate the predicted probabilities $\hat{P}$ for each sample in each category. If the predicted probability $\hat{P}_j(x)$ for a sample $x$ with the label $\tilde{y} = i$ is greater than or equal to the confidence threshold $t_j$, it is considered that the true underlying label $y^*$ for the sample $x$ is $j$ instead of $i$ to some extent. The confidence threshold $t_j$ is obtained by calculating the average predicted probability for samples with the label $\tilde{y} = j$.
\begin{equation}t_{j} : =\frac{1}{X_{\tilde{y}=j } } {\textstyle \sum_{x\in X_{\tilde{y}=j}  }^{}}  \hat{p} _{j} \left ( x \right ) .\label{eq}\end{equation}

Based on the predicted labels, we further introduce the confusion matrix $C_{\tilde{y}, y^*}$, where $C_{\tilde{y}, y^*}[i][j]$ represents the count of samples $x$ with the predicted label $i$ ($\tilde{y} = i$) but the true underlying label may be $j$ ($y^* = j$). In formal terms, $C_{\tilde{y}, y^*}$ can be defined as:
\begin{equation}C_{\tilde{y},y^{\ast }   } \left [ i \right ] \left [ j \right ] : = \left | \hat{X}_{\tilde{y}=i,y^{\ast }=j}   \right | .\label{eq}\end{equation}
\begin{equation}\hat{X}_{\tilde{y}=i,y^{\ast }=j}: = \left \{ x\in X_{\tilde{y}=j }:\hat{p}_{j}\left ( x \right ) \ge t_{j},j=\underset{l\in M:\hat{p}_{l}\left ( x \right ) \ge t_{j}}{ arg max}\hat{p}_{l}\left ( x \right )    \right \} .\label{eq}\end{equation}

By constructing the confusion matrix $C_{\tilde{y}, y^*}$, we can further estimate the $m \times m$ joint distribution matrix $Q_{\tilde{y}, y^*}$ for $p(\tilde{y}, y^*)$:
\begin{equation}Q_{\tilde{y},y^{\ast }   } \left [ i \right ] \left [ j \right ] = \frac{\frac{C_{\tilde{y},y^{\ast }   } \left [ i \right ] \left [ j \right ]}{ {\textstyle \sum_{j\in M}^{}C_{\tilde{y},y^{\ast }   } \left [ i \right ] \left [ j \right ]} }\cdot \left | X_{\tilde{y}=j }  \right |  }{ {\textstyle \sum_{i\in M,j\in M}^{}\left (\frac{C_{\tilde{y},y^{\ast }   } \left [ i \right ] \left [ j \right ]}{ {\textstyle \sum_{j\in M}^{}C_{\tilde{y},y^{\ast }   } \left [ i \right ] \left [ j \right ]} }\cdot \left | X_{\tilde{y}=j }  \right |    \right ) } } .\label{eq}\end{equation}

Next, we utilize the Prune by Class (PBC) method to identify incorrect labels. Specifically, for each class $i \in \mathcal{M}$, PBC selects the samples with the lowest confidence $\hat{p}(\tilde{y} = i; x \in X_i)$ as samples with incorrect annotations. The number of selections is given by:
\begin{equation}n\cdot  {\textstyle \sum_{j\in M,j\ne  i}^{}}\left (Q_{\tilde{y},y^{\ast }   } \left [ i \right ] \left [ j \right ]   \right )  .\label{eq}\end{equation}

After filtering out untrusted samples and considering data missingness, the class weights for each category are adjusted using the following formula:
\begin{equation}\frac{1}{\hat{p} \left (  \tilde{y}=i\mid y^{\ast }=i  \right ) }=\frac{\hat{ Q   } _{y^{\ast }} \left [ i \right ]}{\hat{ Q   }_{\tilde{y},y^{\ast }   } \left [ i \right ] \left [ j \right ]}  .\label{eq}\end{equation}

\begin{equation} Q  _{y^{\ast }   } \left [ i \right ] = {\textstyle \sum_{j\in M}^{}}  Q  _{\tilde{y},y^{\ast }   } \left [ j \right ] \left [ i \right ].\label{eq}\end{equation}

Therefore, high-confidence labels and samples serve as input to the XGBoost classifier, designed as an extensible tree-based algorithm within the machine learning system. Its purpose is to predict the remaining unlabeled pixels in a region, indicating their affiliation with specific microscopic structural categories. In the first stage, pathologists provide pixel-level annotations for cellular, fibrous, and glial structures in color images, which we map onto polarization parameter images. During the training of the pathological tissue structure classifier in the first stage, the input consists of polarization parameters for each pixel, and the output is a probability map for the structural classification of each pixel. Due to the limited and coarse-grained nature of the annotations provided by doctors, and aiming to enhance classification performance, we employ confidence learning and decision tree classifiers to improve the accuracy of structure recognition. In the first stage, we obtain a probability map for each pixel in the measurement area, indicating the likelihood of belonging to a specific structure. This probability map serves as the target for the second-stage network.

\subsubsection{Utilizing Encoder and Decoder to Extract Structural Information}
CNN has been widely applied in image classification tasks. However, when using the pixel-level images (1591× 2291 pixels) generated in the first stage as input, it is challenging to apply CNN directly to the classification task. Firstly, training CNN on a large number of high-resolution images is computationally difficult due to increased input size and model parameter count. Secondly, complex models require abundant samples for training, and the available sample quantity is insufficient for training a sophisticated model. Therefore, there is a need to reduce model complexity, compress embedding dimensions, and simultaneously decrease parameter count and computational load. Additionally, preserving the pixel-level annotations from the first stage as prior knowledge is desired to retain the microscopic structural information predicted in the first stage. Our solution involves dividing polarimetric parameter images into smaller image blocks (224×224 pixels) for direct input to CNN. Subsequently, CNN is employed for feature extraction, converting all polarimetric parameter image blocks into a collection of low-dimensional feature embeddings (Embedding Layer, EMB). After feature extraction, both training and prediction can be performed in the low-dimensional feature space rather than the high-dimensional pixel space. This significantly reduces the computational load required for training supervised deep learning models.

Popular pre-trained network models such as ResNet \cite{35} and pre-activation ResNet \cite{39} have proven effective for various image classification tasks. They offer a universal encoder that encompasses low and high-level feature extraction. Therefore, when designing the second layer network, we leverage the existing feature extraction architecture of mature networks and focus on exploring the optimal approach for training models with polarimetric data. Given that the second stage aims to learn a universal feature representation at the image block level through training a classification network, we adopt the common encoder-decoder classification structure, with the encoder being any basic structure. In this work, we use the ResNet series model as the basic structure for the encoder layer, with minor modifications to enable training with polarimetric data. During training, a simple two-layer convolutional network is employed as the decoder to combine features from the encoder. To calculate the difference between the network output and the structural classification probability map from the first stage, we directly upsample the feature maps from the decoder to the size of the corresponding probability image. This simple decoder design allows the network to focus on training a universal encoder. In the testing phase, the decoder is removed, and the trained encoder can be transferred to the third stage's task. Therefore, the first half of the encoder-decoder structure information extraction network relies on the representation extraction ability of the pre-trained model, maximizing the advantages of the pre-trained model. Meanwhile, the simple decoder design ensures that the information about the microscopic structure provided by the probability map is fed back to the encoder as much as possible. With this design, we aim for the encoder to learn both the correlation between different pixels and distill the information about the microscopic structure learned in the first stage. As a result, the representation EMB learned in the second stage possesses both the information from the first stage's structure and local correlation information.

In the second stage, for each pixel $x_i$, the learned feature representation $\text{EMB}$, after passing through the decoder, is used to regenerate the probability of belonging to a certain structural class, denoted as $\hat{P}_{decoder}(x_i)$. The probability of belonging to a certain structural class predicted by the first stage for each pixel is denoted as $\hat{P}_{XGB}(x_i)$. To validate the consistency between these two prediction results, we use two quantitative metrics: accuracy and loss function. Accuracy measures the ratio of pixels for which the predicted structure class with the highest probability in the first stage matches the one predicted by the decoder, and is defined as:
\begin{equation}acc= \frac{ \sum_{i=1}^{n}1\left ( argmax\hat{p}  _{decoder}\left ( x_{i}  \right )= arg max\hat{p}  _{XGB}\left ( x_{i}  \right ) \right )  }{n} .\label{eq}\end{equation}

Where the indicator function $1(x)$ is defined as:
\begin{equation}1(x) =
\begin{cases}
  1, & \text{if condition $x$ is true} \\
  0, & \text{if condition $x$ is false}
\end{cases}.\label{eq}\end{equation}

The loss function is defined as the difference between the predicted structural labels in the first and second stages, given by:
\begin{equation}loss=\frac{\sum_{i=1}^{n}RMSE\left ( \hat{P}_{decoder}\left ( x_{i},  \right ),\hat{P}_{XGB}\left ( x_{i}  \right )    \right )  }{n} .\label{eq}\end{equation}
The root mean square error (RMSE) is defined as:
\begin{equation}RMSE\left ( Y,\hat{Y}  \right ) =\sqrt{\frac{\sum_{i=1}^{4}\left ( Y_{i}-\hat{Y}_{i}   \right )^{2}   }{4} } .\label{eq}\end{equation}

Where $Y$ is the label predicted by the decoder in the second stage, $Y^*$ is the target label for each pixel, and it is the predicted result label generated by the XGBoost classifier. The target labels for each pixel are not the actual ground truth labels. Due to the scarcity of real pixel-level annotation information, we use the XGBoost classifier pre-trained in the first stage as a pseudo-labeler to provide target labels.

\subsubsection{Applying Attention Mechanism to Classify Pathological Samples}
\begin{figure}[!t]
\centerline{\includegraphics[width=\columnwidth]{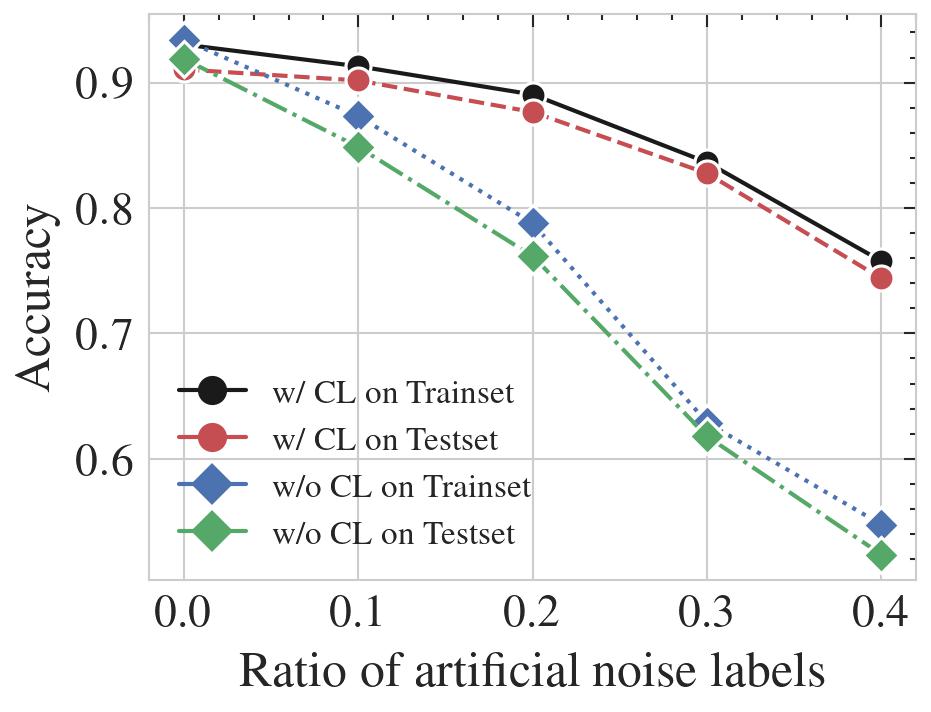}}
\caption{Polarization parameter maps for different slice thicknesses. (a) 4$\upmu$m thick H\&E stained slice. (b) 4$\upmu$m thick unstained slice. (c) 6$\upmu$m thick unstained slice. (d) 8$\upmu$m thick unstained slice. (e) 10$\upmu$m thick unstained slice. (f) 12$\upmu$m thick unstained slice.}
\label{fig3}
\end{figure}
\begin{figure*}[!t]
\centerline{\includegraphics[width=\linewidth]{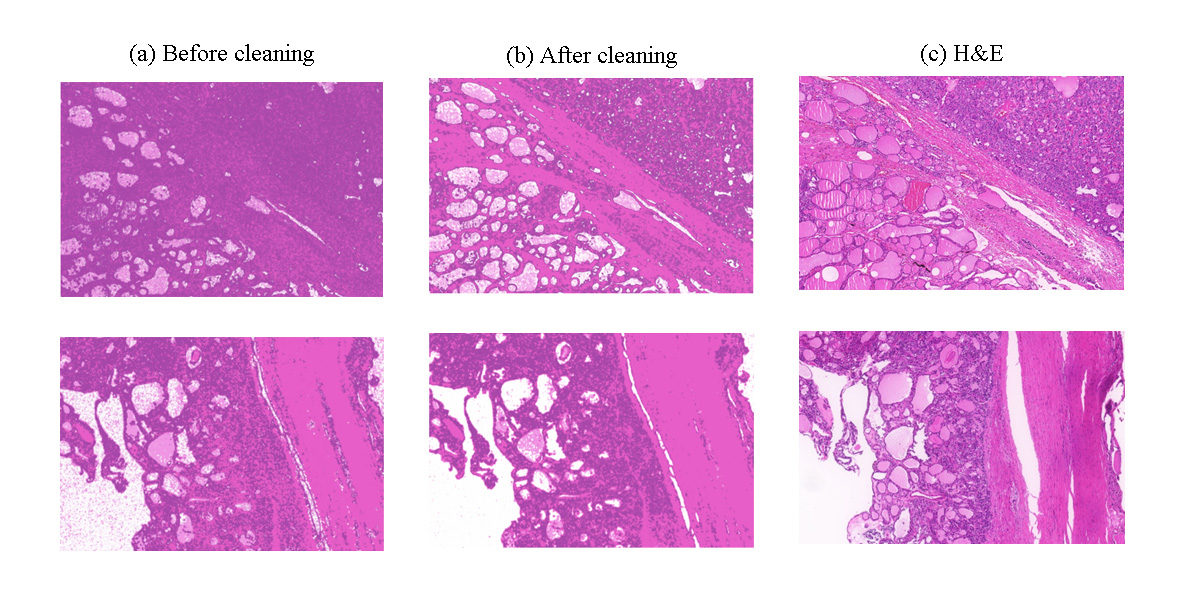}}
\caption{ Comparison of Microstructure Recognition Results. (a) Without Confidence Learning. (b) With Confidence Learning. (c) Corresponding H\&E Image.}
\label{fig4}
\end{figure*}

In addition to using max pooling, other operators such as average operator, weighted average operator, and quantile truncation can be employed to filter or aggregate representations of different image blocks. However, their limitations lie in limited flexibility and sensitivity to data, often requiring tedious fine-tuning for specific datasets. To provide a simple, intuitive mechanism that is both flexible and user-friendly for aggregating block-level representations with good model interpretability, we designed a trainable attention-based pooling function to aggregate block-level representations and obtain ROI-level representations for downstream classification tasks.

In our designed multi-class attention pooling network, the downstream attention network first obtains ROI-level representations based on the attention mechanism, and then trains three binary classifiers simultaneously based on these representations to obtain expected scores for the three types of representations.

The specific implementation details are as follows: we use two fully connected networks to remap and project the block-level representations $V_a$ and $U_a$, projecting them onto a dense 256-dimensional latent space. We then obtain attention scores for each block representation through softmax operations. Here, assuming the operation is on the $i$-th ROI region, which is divided into a total of $K$ blocks, and the representation of the $k$-th block is denoted as $e_{i,k}$ (obtained from the output of the encoder layer of the pre-trained model in the second stage), the shared two projection functions are: $V_a \in \mathbb{R}^{256 \times 1024}$ with a $\tanh$ activation function, and $U_a \in \mathbb{R}^{256 \times 1024}$ with a sigmoid activation function (denoted as $\text{sigm}$). Dropout layers ($P=0.25$) are added after both layers as regularization terms to prevent overfitting. Thus, we can calculate the attention weight for the $k$-th patch as follows:
\begin{equation}a_{i,k}=\frac{\exp \left \{ W_{a,i}\left ( \tanh \left ( V_{a}e_{i,k} \right ) \odot sigm\left ( U_{a}e_{i,k} \right )  \right )  \right \}  }{\sum_{j=1}^{k}\exp \left \{ W_{a,i}\left ( \tanh \left ( V_{a}e_{i,k} \right ) \odot sigm\left ( U_{a}e_{i,k} \right )  \right )  \right \} }  .\label{eq}\end{equation}

Here,  $\bigodot $ represents the inner product operator. Furthermore, aggregating block representations based on attention scores yields the aggregated ROI-level representation as follows:
\begin{equation}e_{roi,i}= {\textstyle \sum_{k=1}^{K}a_{i,k}e_{i,k}} .\label{eq}\end{equation}

This type of aggregation allows adaptive learning of weights based on block representations and maintains interpretability, where regions with higher weights are more likely to be critical for diagnostic decisions. Downstream of learning the structured aggregation representation with the attention network, we simultaneously train three classification layers  $W_1, W_2, W_3 \in \mathbb{R}^{1 \times 1024}$ , along with sigmoid activation functions, to obtain three binary prediction scores. These scores can be further transformed using the softmax function to obtain the final phenotype prediction at the ROI level.

\section{Results and Discussion}

\subsection{Pathological Structure Recognition Results and Analysis of Polarization Feature Importance}

To verify the feasibility of confidence learning, we randomly shuffled and artificially added noise to a portion of the annotations provided by pathologists. Under different proportions of artificial noisy labels, the proposed pathological structure recognition method was used to predict pixels in unlabeled regions of the training and test sets, with accuracy selected as the evaluation metric. As shown in Fig. 3, we calculated the pathological structure recognition results for four scenarios: black dots represent accuracy with confidence learning added in the training set, red dots represent accuracy with confidence learning added in the test set, blue dots represent accuracy without confidence learning in the training set, and green dots represent accuracy without confidence learning in the test set.
\begin{figure}[!t]
\centerline{\includegraphics[width=\columnwidth]{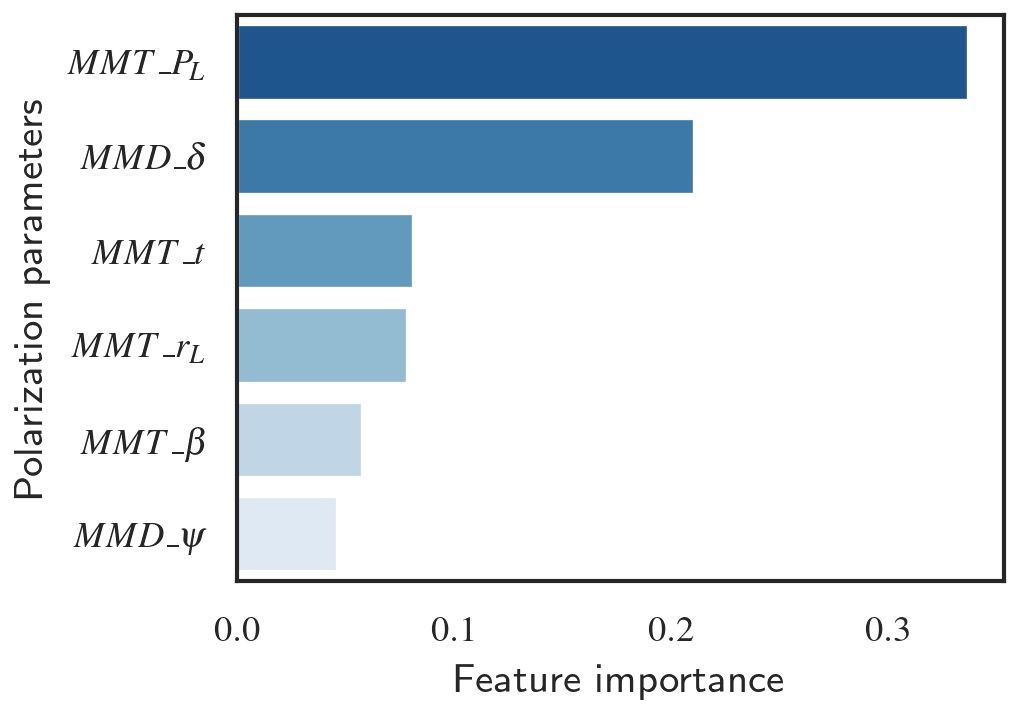}}
\caption{Importance of Polarization Features. Demonstrates the pivotal role played by polarization features in the process of microstructure recognition.}
\label{fig5}
\end{figure}
From the recognition accuracy, it can be observed that the accuracy with confidence learning added is higher than the accuracy without confidence learning, both in the training and test sets. As the proportion of noisy labels increases, the accuracy drop with confidence learning is slower than the drop without confidence learning, indicating that the model's recognition ability remains relatively stable. The accuracy on the training set can to some extent represent the training effect of the model, while the accuracy on the test set can represent the model's transfer and generalization ability. This demonstrates that the proposed pathological structure recognition method can improve classification performance.

Therefore, confidence learning can enhance the accuracy of predictions in unlabeled regions. By using a small amount of labeled data with confidence learning to train the classifier within a region, the trained classifier can predict the structures in the remaining region, increasing the credibility of the prediction results. Moreover, if there are mislabeled instances in the initial training annotations, this situation may lead to incorrect classifier predictions. The introduction of confidence learning can correct some of the classification results, reducing the impact of labeling issues.

To visually demonstrate the comparison between recognition results with and without confidence learning, we present the three recognition outcomes in the familiar H\&E color scheme for pathologists, as shown in Fig. 4. It can be observed that the addition of confidence learning leads to more confident identification of some originally uncertain pixels, manifested in clearer colors for certain pixels. Additionally, some initially misidentified pixels are corrected to the right identification, as evident in color differences before and after introducing confidence learning. This representation in polarized virtual staining emulates the commonly used pathology observation format, making it convenient for pathologists to observe pathological structures. By comparing with adjacent stained H\&E images, it can be seen that the recognition results for pathological structures are reasonably accurate, demonstrating the feasibility of this recognition method.

\begin{figure}[!t]
\centerline{\includegraphics[width=\columnwidth]{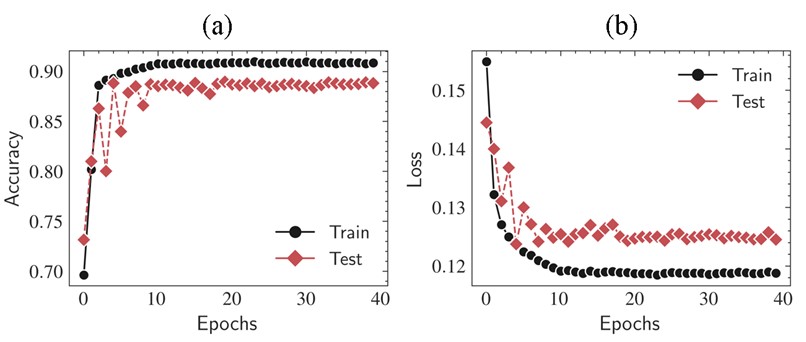}}
\caption{Curves of Accuracy and Loss Function Metrics. (a) Accuracy. (b) Loss Function.}
\label{fig6}
\end{figure}

Next, we conducted an analysis of the feature importance of different polarization parameters in the process of microstructure recognition. This analysis helps us understand the importance of each polarization parameter in the model's prediction process. Since polarization parameters have clear physical meanings, we can interpret the differences in optical features during microstructure recognition by analyzing the physical significance of polarization parameters that are highly relevant to the classification task.

As shown in Fig. 5, in the process of pathological structure recognition in thyroid tumors, the following polarization parameters are considered important, providing insights into the optical properties that play a significant role in the recognition process from an interpretable perspective: linear retardance, linear phase delay, degree of anisotropy, ability to convert circular polarization to linear polarization, circular birefringence, and circular phase delay.
\begin{figure}[!t]
\centerline{\includegraphics[width=\columnwidth]{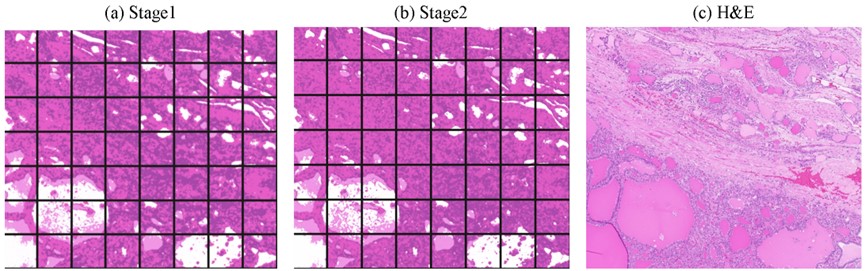}}
\caption{Structure Classification Probability Maps. (a) Classification results from the first stage. (b) Reconstruction results from the second stage. (c) Corresponding H\&E image.}
\label{fig7}
\end{figure}
\subsection{Evaluation of Structural Information Recovery Performance}
The results of the accuracy and loss metrics in the second stage are shown in Fig. 6. It can be observed that due to the presence of a pre-trained model, the model converges quickly (approaching convergence at epoch=10), demonstrating the advantage of introducing pre-trained models. Transferring prior knowledge pre-trained on a common image dataset significantly accelerates the training speed and reduces the need for a large number of samples. As training progresses, the loss metric continuously decreases, and the accuracy metric continuously increases, indicating that the second stage can adequately distill information from the first stage, showing a high consistency between the two in predicting structural information.

\begin{figure*}[!t]
\centerline{\includegraphics[width=\linewidth]{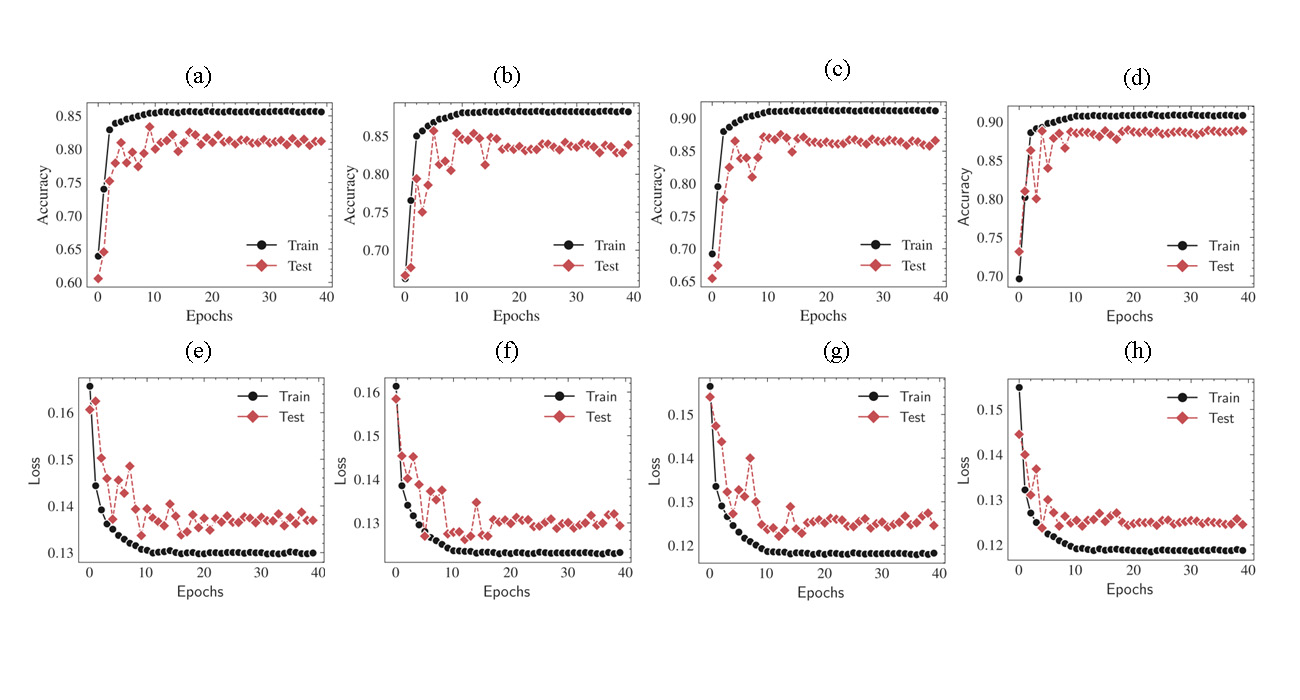}}
\caption{Loss curves using different models as encoders. (a) Accuracy of ResNet18. (b) Accuracy of ResNet34. (c) Accuracy of ResNet50. (d) Accuracy of ResNet101. (e) Loss of ResNet18. (f) Loss of ResNet34. (g) Loss of ResNet50. (h) Loss of ResNet101. }
\label{fig8}
\end{figure*}

The comparison of the structural classification probability maps between the two stages is shown in Fig. 7. It can be observed that the second stage can highly reconstruct the structural information from the first stage. The reconstructed result can serve as an approximation of the pathology color image. In summary, the second stage can leverage the pre-trained model and convolutional network structure to extract correlations between different pixels. The learned feature representations, combined with the decoder, can achieve pixel-level structural information prediction. The learned feature representations have two key properties: first, benefiting from the presence of pre-training, they capture inter-pixel correlations and leverage prior knowledge pre-trained on another dataset in the image domain; second, they possess the ability to predict pixel-level structural information.

Moreover, to select the most suitable ResNet model, we analyzed different models as encoders for the second stage based on the accuracy and loss metrics. From Fig. 8, we can observe that using pre-trained models leads to faster convergence, with models approaching convergence at around epochs=10. Even with only one training epoch (epochs=0), the accuracy already exceeds 0.6, demonstrating the superiority of using pre-trained models. When choosing a smaller pre-trained model, the final prediction performance may slightly decrease, indicating that smaller models learn representations of lower quality. As the model size increases, the prediction performance improves. On the other hand, we can observe that the performance difference between ResNet models on the training set and the test set is small. This suggests that larger pre-trained models may be less prone to overfitting during fine-tuning, indicating better generalization. Additionally, comparing ResNet101 and ResNet50, we find that the performance difference between the two models is already small. Further increasing the model size may not yield significant gains. Therefore, in our experiments, we chose ResNet101.

\subsection{Evaluation of Classification Performance }
\begin{figure}[!t]
\centerline{\includegraphics[width=\columnwidth]{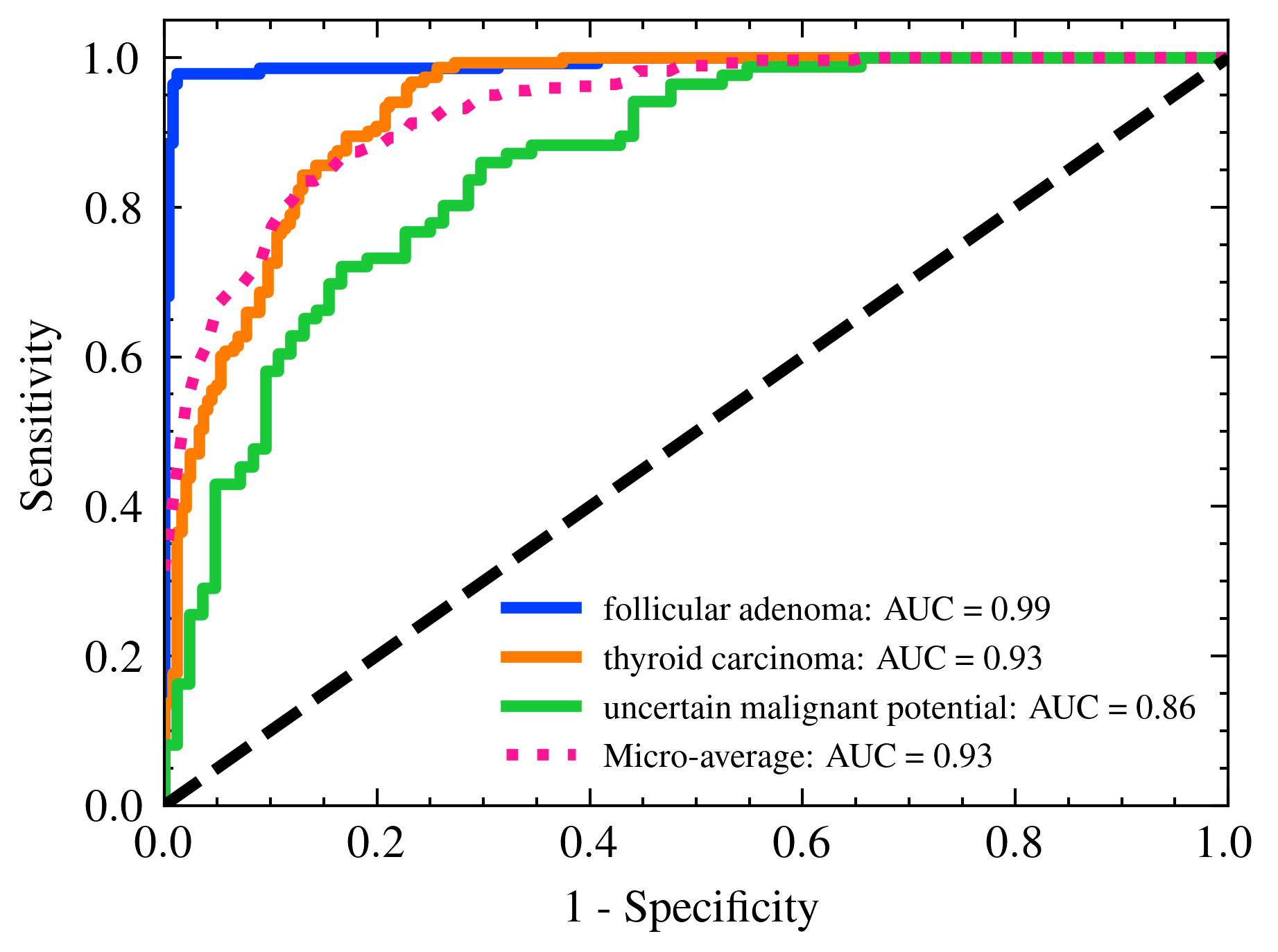}}
\caption{ROC curves and corresponding AUC metrics.}
\label{fig9}
\end{figure}
\begin{figure*}[!t]
\centerline{\includegraphics[width=\linewidth]{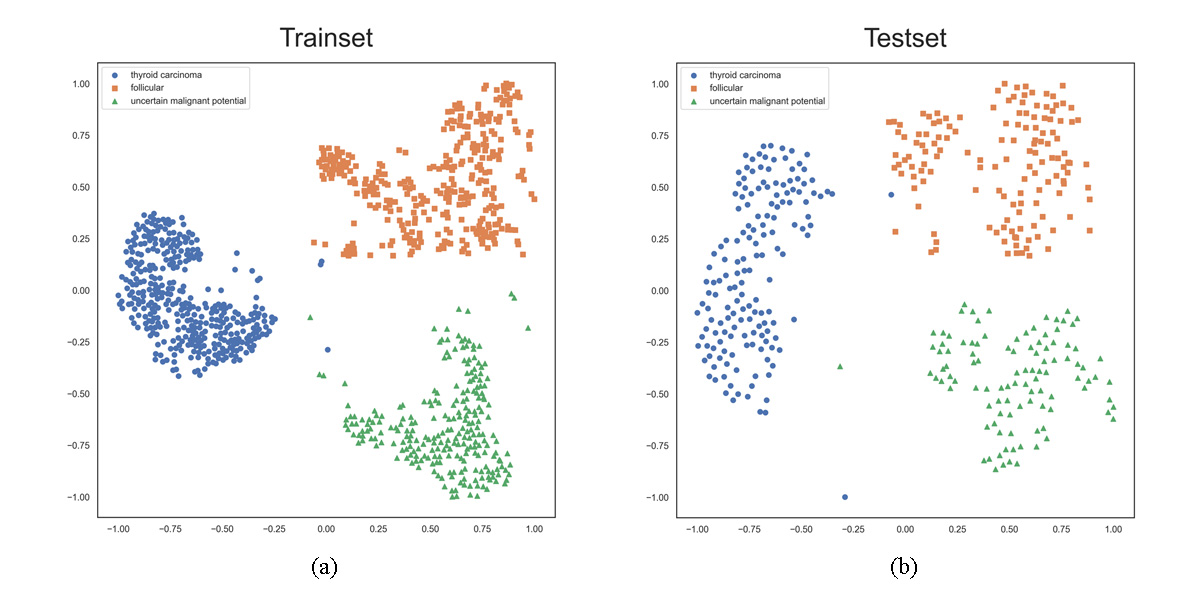}}
\caption{Visualization of the classification of ROI-level feature representation. (a) Classification visualization of three tumors on the training set. (b) Classification visualization of three tumors on the test set.}
\label{fig10}
\end{figure*}
In this work, we selected the Area Under the ROC Curve (AUC) and Micro-average AUC as metrics to demonstrate the effectiveness of our proposed model. AUC is a value between 0 and 1, measuring the classifier's ability to distinguish between positive and negative instances. A higher AUC indicates better classifier performance. Micro-average AUC consolidates the prediction results of all samples and calculates the AUC value for the entire dataset, providing a more comprehensive reflection of the classifier's overall performance. 

As shown in Fig. 9, the ROC curves and corresponding AUC values for the proposed network in the classification of thyroid tumors demonstrate good discrimination among malignant thyroid tumors, benign tumors, and indeterminate borderline tumors. Particularly, the classification performance is excellent for malignant thyroid tumors with an AUC of 0.99. The discrimination for benign tumors follows with an AUC of 0.93. Although the discrimination for indeterminate borderline tumors shows a slightly lower AUC compared to the previous two categories, it remains substantial at 0.86. Moreover, the Micro-average AUC is 0.93, confirming the effective differentiation of these three types of thyroid tumors using polarimetric features.

In the third stage, utilizing the attention mechanism to learn the weights of different image blocks, we aggregate all the image block embeddings (EMB) into the entire Region of Interest (ROI)'s EMB. Subsequently, we introduce an LR classifier based on the ROI's EMB for ROI-level classification tasks. The final learned ROI's EMB is visualized in Fig. 10, demonstrating its classification capability. The aggregated EMBs exhibit distinctiveness among ROIs, confirming the effectiveness of the proposed attention mechanism in integrating EMBs from different image blocks. The meaningfulness of the learned weights for various image blocks is also validated through this approach.

\section{Conclusion}
This study focuses on polarization-modality-based representation learning for pathological diagnosis of thyroid tumors. Difficulty in pathological diagnosis of thyroid tumors arises from inadequate sampling, and accurately distinguishing between benign and malignant thyroid tumors is crucial for devising optimal diagnosis and treatment plans in clinical settings. We designed a three-stage classification model based on polarization-modality representation learning. The network comprises a pathology structure recognition method, an encoder-decoder structure for extracting pixel-level annotation information, and an attention-based learning mechanism for the final classification task. Given the limited and coarse-grained pixel-level annotations provided by pathologists, we enhance microstructure recognition accuracy using confidence learning and a decision tree classifier to obtain a probability map of the structure category for each pixel in the measurement region, serving as the target for the second-stage encoder and decoder. Leveraging the representation extraction capability of pre-trained models and a simple decoding design, the encoder learns the correlation between different pixels and distills pixel-level structure annotation information. Region prediction is improved by integrating local information through importance learning. Additionally, this study transfers the method of extracting polarization feature parameters from stained samples to unstained samples. Polarization features representing microstructure are extracted for the measured pathological region, and polarization pseudo-coloring is applied to characterize the thyroid target region. The synthesized polarization pseudo-color images are closely related to pathological diagnosis and can be used by pathologists for diagnosis on unstained samples, reducing the need for stained pathological slides. Experimental results demonstrate that, by leveraging the concept of indirect measurement and extracting polarization features related to lesions, our method is capable of determining sample types in cases where the capsule structure cannot be obtained from the specimen. This research aims to provide an objective and highly accurate indirect diagnostic technique for thyroid tumors.

\bibliographystyle{IEEEtran}
\bibliography{ref.bib}

\begin{thebibliography}{10}
\providecommand{\url}[1]{#1}
\csname url@samestyle\endcsname
\providecommand{\newblock}{\relax}
\providecommand{\bibinfo}[2]{#2}
\providecommand{\BIBentrySTDinterwordspacing}{\spaceskip=0pt\relax}
\providecommand{\BIBentryALTinterwordstretchfactor}{4}
\providecommand{\BIBentryALTinterwordspacing}{\spaceskip=\fontdimen2\font plus
\BIBentryALTinterwordstretchfactor\fontdimen3\font minus
  \fontdimen4\font\relax}
\providecommand{\BIBforeignlanguage}[2]{{%
\expandafter\ifx\csname l@#1\endcsname\relax
\typeout{** WARNING: IEEEtran.bst: No hyphenation pattern has been}%
\typeout{** loaded for the language `#1'. Using the pattern for}%
\typeout{** the default language instead.}%
\else
\language=\csname l@#1\endcsname
\fi
#2}}
\providecommand{\BIBdecl}{\relax}
\BIBdecl

\bibitem{1}
M.~Luster, L.~H. Duntas, and L.~Wartofsky, \emph{The thyroid and its
  diseases}.\hskip 1em plus 0.5em minus 0.4em\relax Springer, 2019.

\bibitem{2}
B.~S. Chhikara and K.~Parang, ``Global cancer statistics 2022: the trends
  projection analysis,'' \emph{Chemical Biology Letters}, vol.~10, no.~1, pp.
  451--451, 2023.

\bibitem{3}
Z.~W. Baloch, S.~L. Asa, J.~A. Barletta, R.~A. Ghossein, C.~C. Juhlin, C.~K.
  Jung, V.~A. LiVolsi, M.~G. Papotti, M.~Sobrinho-Simoes, G.~Tallini
  \emph{et~al.}, ``Overview of the 2022 who classification of thyroid
  neoplasms,'' \emph{Endocrine pathology}, vol.~33, no.~1, pp. 27--63, 2022.

\bibitem{4}
M.~Wr{\'o}blewski, J.~Wr{\'o}blewska, J.~Nuszkiewicz, M.~Paw{\l}owska,
  R.~Weso{\l}owski, and A.~Wo{\'z}niak, ``The role of selected trace elements
  in oxidoreductive homeostasis in patients with thyroid diseases,''
  \emph{International Journal of Molecular Sciences}, vol.~24, no.~5, p. 4840,
  2023.

\bibitem{5}
Q.~T. Nguyen, E.~J. Lee, M.~G. Huang, Y.~I. Park, A.~Khullar, and R.~A.
  Plodkowski, ``Diagnosis and treatment of patients with thyroid cancer,''
  \emph{American health \& drug benefits}, vol.~8, no.~1, p.~30, 2015.

\bibitem{6}
D.~W. Chen, B.~H. Lang, D.~S. McLeod, K.~Newbold, and M.~R. Haymart, ``Thyroid
  cancer,'' \emph{The Lancet}, vol. 401, no. 10387, pp. 1531--1544, 2023.

\bibitem{7}
M.~E. Cabanillas, D.~G. McFadden, and C.~Durante, ``Thyroid cancer,'' \emph{The
  Lancet}, vol. 388, no. 10061, pp. 2783--2795, 2016.

\bibitem{8}
C.~M. Kitahara and J.~A. Sosa, ``The changing incidence of thyroid cancer,''
  \emph{Nature Reviews Endocrinology}, vol.~12, no.~11, pp. 646--653, 2016.

\bibitem{9}
Y.~H. Sharifovna, ``Thyroid cancer diagnostics, classification, staging,''
  \emph{Ijtimoiy fanlarda innovasiya onlayn ilmiy jurnali}, vol.~1, no.~5, pp.
  63--69, 2021.

\bibitem{10}
M.~Schlumberger and S.~Leboulleux, ``Current practice in patients with
  differentiated thyroid cancer,'' \emph{Nature Reviews Endocrinology},
  vol.~17, no.~3, pp. 176--188, 2021.

\bibitem{11}
M.~Cui and D.~Y. Zhang, ``Artificial intelligence and computational
  pathology,'' \emph{Laboratory Investigation}, vol. 101, no.~4, pp. 412--422,
  2021.

\bibitem{12}
B.~Acs, M.~Rantalainen, and J.~Hartman, ``Artificial intelligence as the next
  step towards precision pathology,'' \emph{Journal of internal medicine}, vol.
  288, no.~1, pp. 62--81, 2020.

\bibitem{13}
K.~Bera, K.~A. Schalper, D.~L. Rimm, V.~Velcheti, and A.~Madabhushi,
  ``Artificial intelligence in digital pathology—new tools for diagnosis and
  precision oncology,'' \emph{Nature reviews Clinical oncology}, vol.~16,
  no.~11, pp. 703--715, 2019.

\bibitem{14}
A.~H. Song, G.~Jaume, D.~F. Williamson, M.~Y. Lu, A.~Vaidya, T.~R. Miller, and
  F.~Mahmood, ``Artificial intelligence for digital and computational
  pathology,'' \emph{Nature Reviews Bioengineering}, vol.~1, no.~12, pp.
  930--949, 2023.

\bibitem{dong2023polarization}
J.~Dong, Y.~Yao, L.~Lin, Y.~Dong, J.~Wan, R.~Peng, C.~Li, and H.~Ma, ``A
  polarization and radiomics feature fusion network for the classification of
  hepatocellular carcinoma and intrahepatic cholangiocarcinoma,'' \emph{arXiv
  preprint arXiv:2312.16607}, 2023.

\bibitem{15}
F.~Ghaznavi, A.~Evans, A.~Madabhushi, and M.~Feldman, ``Digital imaging in
  pathology: whole-slide imaging and beyond,'' \emph{Annual Review of
  Pathology: Mechanisms of Disease}, vol.~8, pp. 331--359, 2013.

\bibitem{16}
C.~D. Gatenbee, A.-M. Baker, S.~Prabhakaran, O.~Swinyard, R.~J. Slebos,
  G.~Mandal, E.~Mulholland, N.~Andor, A.~Marusyk, S.~Leedham \emph{et~al.},
  ``Virtual alignment of pathology image series for multi-gigapixel whole slide
  images,'' \emph{Nature communications}, vol.~14, no.~1, p. 4502, 2023.

\bibitem{17}
C.-W. Wang, Y.-C. Lee, E.~Calista, F.~Zhou, H.~Zhu, R.~Suzuki, D.~Komura,
  S.~Ishikawa, and S.-P. Cheng, ``A benchmark for comparing precision medicine
  methods in thyroid cancer diagnosis using tissue microarrays,''
  \emph{Bioinformatics}, vol.~34, no.~10, pp. 1767--1773, 2018.

\bibitem{18}
D.~Dov, S.~Z. Kovalsky, J.~Cohen, D.~E. Range, R.~Henao, and L.~Carin,
  ``Thyroid cancer malignancy prediction from whole slide cytopathology
  images,'' in \emph{Machine Learning for Healthcare Conference}.\hskip 1em
  plus 0.5em minus 0.4em\relax PMLR, 2019, pp. 553--570.

\bibitem{19}
D.~D. Elliott~Range, D.~Dov, S.~Z. Kovalsky, R.~Henao, L.~Carin, and J.~Cohen,
  ``Application of a machine learning algorithm to predict malignancy in
  thyroid cytopathology,'' \emph{Cancer cytopathology}, vol. 128, no.~4, pp.
  287--295, 2020.

\bibitem{20}
A.~S. El-Hossiny, W.~Al-Atabany, O.~Hassan, A.~M. Soliman, and S.~A. Sami,
  ``classification of thyroid carcinoma in whole slide images using cascaded
  cnn,'' \emph{IEEE Access}, vol.~9, pp. 88\,429--88\,438, 2021.

\bibitem{21}
A.~Gupta, P.~J. Harrison, H.~Wieslander, N.~Pielawski, K.~Kartasalo, G.~Partel,
  L.~Solorzano, A.~Suveer, A.~H. Klemm, O.~Spjuth \emph{et~al.}, ``Deep
  learning in image cytometry: a review,'' \emph{Cytometry Part A}, vol.~95,
  no.~4, pp. 366--380, 2019.

\bibitem{22}
H.~Wieslander, G.~Forslid, E.~Bengtsson, C.~Wahlby, J.-M. Hirsch,
  C.~Runow~Stark, and S.~Kecheril~Sadanandan, ``Deep convolutional neural
  networks for detecting cellular changes due to malignancy,'' in
  \emph{Proceedings of the IEEE international conference on computer vision
  workshops}, 2017, pp. 82--89.

\bibitem{23}
E.~Du, H.~He, N.~Zeng, M.~Sun, Y.~Guo, J.~Wu, S.~Liu, and H.~Ma, ``Mueller
  matrix polarimetry for differentiating characteristic features of cancerous
  tissues,'' \emph{Journal of biomedical optics}, vol.~19, no.~7, pp.
  076\,013--076\,013, 2014.

\bibitem{24}
C.~He, H.~He, J.~Chang, B.~Chen, H.~Ma, and M.~J. Booth, ``Polarisation optics
  for biomedical and clinical applications: a review,'' \emph{Light: Science \&
  Applications}, vol.~10, no.~1, p. 194, 2021.

\bibitem{25}
N.~Ghosh and I.~A. Vitkin, ``Tissue polarimetry: concepts, challenges,
  applications, and outlook,'' \emph{Journal of biomedical optics}, vol.~16,
  no.~11, pp. 110\,801--110\,801, 2011.

\bibitem{26}
H.~He, R.~Liao, N.~Zeng, P.~Li, Z.~Chen, X.~Liu, and H.~Ma, ``Mueller matrix
  polarimetry—an emerging new tool for characterizing the microstructural
  feature of complex biological specimen,'' \emph{Journal of Lightwave
  Technology}, vol.~37, no.~11, pp. 2534--2548, 2018.

\bibitem{27}
Y.~Dong, J.~Wan, L.~Si, Y.~Meng, Y.~Dong, S.~Liu, H.~He, and H.~Ma, ``Deriving
  polarimetry feature parameters to characterize microstructural features in
  histological sections of breast tissues,'' \emph{IEEE Transactions on
  Biomedical Engineering}, vol.~68, no.~3, pp. 881--892, 2020.

\bibitem{28}
Y.~Liu, Y.~Dong, L.~Si, R.~Meng, Y.~Dong, and H.~Ma, ``Comparison between image
  texture and polarization features in histopathology,'' \emph{Biomedical
  Optics Express}, vol.~12, no.~3, pp. 1593--1608, 2021.

\bibitem{29}
K.~Sindhoora, K.~Spandana, D.~Ivanov, E.~Borisova, U.~Raghavendra, S.~Rai,
  S.~Kabekkodu, K.~Mahato, and N.~Mazumder, ``Machine-learning-based
  classification of stokes-mueller polarization images for tissue
  characterization,'' in \emph{Journal of Physics: Conference Series}, vol.
  1859, no.~1.\hskip 1em plus 0.5em minus 0.4em\relax IOP Publishing, 2021, p.
  012045.

\bibitem{30}
X.~Zhou, L.~Ma, W.~Brown, J.~V. Little, A.~Y. Chen, L.~L. Myers, B.~D. Sumer,
  and B.~Fei, ``Automatic detection of head and neck squamous cell carcinoma on
  pathologic slides using polarized hyperspectral imaging and machine
  learning,'' in \emph{Medical Imaging 2021: Digital Pathology}, vol.
  11603.\hskip 1em plus 0.5em minus 0.4em\relax SPIE, 2021, pp. 165--173.

\bibitem{31}
Y.~Dong, J.~Wan, X.~Wang, J.-H. Xue, J.~Zou, H.~He, P.~Li, A.~Hou, and H.~Ma,
  ``A polarization-imaging-based machine learning framework for quantitative
  pathological diagnosis of cervical precancerous lesions,'' \emph{IEEE
  Transactions on Medical Imaging}, vol.~40, no.~12, pp. 3728--3738, 2021.

\bibitem{32}
Y.~Chen, Y.~Dong, L.~Si, W.~Yang, S.~Du, X.~Tian, C.~Li, Q.~Liao, and H.~Ma,
  ``Dual polarization modality fusion network for assisting pathological
  diagnosis,'' \emph{IEEE Transactions on Medical Imaging}, vol.~42, no.~1, pp.
  304--316, 2022.

\bibitem{33}
Y.~Yao, M.~Zuo, Y.~Dong, L.~Shi, Y.~Zhu, L.~Si, X.~Ye, and H.~Ma,
  ``Polarization imaging feature characterization of different endometrium
  phases by machine learning,'' \emph{OSA Continuum}, vol.~4, no.~6, pp.
  1776--1791, 2021.

\bibitem{34}
A.~Krizhevsky, I.~Sutskever, and G.~E. Hinton, ``Imagenet classification with
  deep convolutional neural networks,'' \emph{Advances in neural information
  processing systems}, vol.~25, 2012.

\bibitem{35}
K.~He, X.~Zhang, S.~Ren, and J.~Sun, ``Deep residual learning for image
  recognition,'' in \emph{Proceedings of the IEEE conference on computer vision
  and pattern recognition}, 2016, pp. 770--778.

\bibitem{36}
Y.~Zhao, M.~Reda, K.~Feng, P.~Zhang, G.~Cheng, Z.~Ren, S.~G. Kong, S.~Su,
  H.~Huang, and J.~Zang, ``Detecting giant cell tumor of bone lesions using
  mueller matrix polarization microscopic imaging and multi-parameters fusion
  network,'' \emph{IEEE Sensors Journal}, vol.~20, no.~13, pp. 7208--7215,
  2020.

\bibitem{37}
Y.~Bengio, A.~Courville, and P.~Vincent, ``Representation learning: A review
  and new perspectives,'' \emph{IEEE transactions on pattern analysis and
  machine intelligence}, vol.~35, no.~8, pp. 1798--1828, 2013.

\bibitem{38}
C.~Northcutt, L.~Jiang, and I.~Chuang, ``Confident learning: Estimating
  uncertainty in dataset labels,'' \emph{Journal of Artificial Intelligence
  Research}, vol.~70, pp. 1373--1411, 2021.

\bibitem{39}
K.~He, X.~Zhang, S.~Ren, and J.~Sun, ``Identity mappings in deep residual
  networks,'' in \emph{Computer Vision--ECCV 2016: 14th European Conference,
  Amsterdam, The Netherlands, October 11--14, 2016, Proceedings, Part IV
  14}.\hskip 1em plus 0.5em minus 0.4em\relax Springer, 2016, pp. 630--645.

\end{thebibliography}
\end{document}